\documentclass[onecolumn,amsmath,amssymb,superscriptaddress,nofootinbib,11pt,a4paper]{revtex4}
\usepackage{titlesec}
\titleformat{\section}
  {\normalfont\Large\bfseries}
  {\thesection}
  {1em}
  {}
\titleformat{\subsection}
  {\normalfont\large\bfseries}
  {\thesubsection}
  {1em}
  {}
\pdfoutput=1
\usepackage[T1]{fontenc}
\usepackage{xcolor}
\usepackage{amsfonts}
\usepackage{epstopdf}
\usepackage{wrapfig}
\usepackage{subcaption}
\usepackage{graphicx}
\usepackage{dcolumn}
\usepackage{bm}
\usepackage{float}
\usepackage[utf8]{inputenc}
\usepackage{booktabs}
\usepackage{orcidlink}
\usepackage{multirow}

\begin{document}

\title{Energy extraction from Kerr--Bertotti--Robinson black holes: polar region collapse and the capture of negative energy protons}

\author{Ke Wang\orcidlink{0009-0009-4723-3403}}
\affiliation{College of Physics and Optoelectronic Engineering, Chongqing Normal University, \\Chongqing 401331, China}
\affiliation{School of Material Science and Engineering, Chongqing Jiaotong University, Chongqing 400074, China}

\author{Xiao-Xiong Zeng\orcidlink{0000-0002-2145-9334}}
\email{xxzengphysics@163.com}
\affiliation{College of Physics and Optoelectronic Engineering, Chongqing Normal University, \\Chongqing 401331, China}

\begin{abstract}
The Kerr–Bertotti–Robinson black hole is an exact Einstein–Maxwell solution describing a rotating black hole in an asymptotically uniform magnetic field. It differs from the Kerr black hole in the Wald solution: the magnetic field lines are no longer cylindrical, with the constant-\(A_\phi\) lines saturating at large distances, and the interface \(\vec{E}\cdot\vec{B}=0\) is no longer a cone but contracts with radius and eventually disappears. Determining the acceleration of charged particles pointwise, we find that protons are driven outward and electrons inward in the polar region, and vice versa in the equatorial region; the polar region shrinks as the magnetic field grows. Captured protons carry negative energy and angular momentum, whereas captured electrons carry positive energy and angular momentum. Fixing the capture domain by the gauge-invariant flux criterion and integrating over the \((r,\theta)\) plane, we find that in weak magnetic fields the black hole gains mass, while in strong magnetic fields energy extraction occurs. Energy extraction can be achieved under the usual spherically symmetric density, without introducing the anisotropic density distribution required in the Wald solution for a Kerr black hole.
\end{abstract}
\maketitle

\newpage
\section{Introduction}
Black holes are among the most striking predictions of general relativity, and the interaction between the extreme gravitational field and strong electromagnetic field around them has long been a central topic in relativistic astrophysics. In recent years, imaging observations of M87* and Sgr A* by the Event Horizon Telescope have for the first time revealed direct evidence for black hole shadows and the magnetic field structure near them\cite{7,8,9,10}, while gravitational wave detections have continuously deepened our understanding of black hole mergers and spacetime properties\cite{11}. In these observations, magnetic fields play a crucial role: jets from active galactic nuclei\cite{12,13}, high-energy emission from gamma-ray bursts\cite{14,15}, and matter transport in black hole accretion disks\cite{16,17} are all believed to be closely related to strong magnetic fields around black holes. Therefore, constructing reasonable magnetized black hole models and studying their electrodynamic processes has important theoretical and observational significance for explaining these high-energy astrophysical phenomena.

On the theoretical side, exact solutions describing black holes immersed in external magnetic fields have always been an important problem at the intersection of general relativity and electromagnetism. The earliest and most widely used model is the Wald solution\cite{18}, which describes a Kerr black hole immersed in an asymptotically uniform test magnetic field aligned with the rotation axis. The Wald solution is very effective in the weak-field approximation and reveals the twisting of magnetic field lines near the horizon and the existence of an induced electric field. Reference \cite{5} studied the electrodynamic processes of a Kerr black hole in the Wald solution. They concluded that in the polar region, the electric field accelerates electrons outward and protons inward; in the equatorial region, it accelerates electrons inward and protons outward. The division between the polar and equatorial regions is a fixed value independent of the magnetic field. They also analyzed the capture of electrons and protons by the black hole and found that outside the ergosphere, with a spherically symmetric density, the black hole mass increases, but with an anisotropic density, energy extraction can occur. Reference \cite{6} extended the study of \cite{5} by analyzing the electrodynamic processes of a Kerr black hole in the Wald solution with test charges, in this case, with a spherically symmetric density and special charge values, energy extraction can occur.

The energy extraction achieved by the above mechanism is essentially different from the classical Penrose process\cite{64}. The Penrose process requires a particle to first enter the ergosphere and split into two parts inside it. One part falls into the black hole along a negative-energy orbit, while the other escapes to infinity carrying more energy than the original particle. Its validity relies on two ingredients. The first is the existence of an ergosphere, hence requiring black hole rotation, and the second is the splitting of particles inside the ergosphere. Its negative energy is purely gravitational. Inside the ergosphere $g_{tt}>0$, so that timelike orbits can still have $E<0$. In the electrodynamic processes studied in Refs. \cite{5,6}, the captured particles are initially at rest outside the ergosphere. The particles neither need to enter the ergosphere nor need to split. Their negative energy comes entirely from the electromagnetic coupling term $-q_iA_t$. When the electromagnetic term dominates, even if the particle is outside the ergosphere, where $g_{tt}<0$ and a neutral particle at rest must have $E=m\sqrt{-g_{tt}}>0$, as long as $q_iA_t>0$, one can still obtain $E_i<0$. Therefore, this is a single-particle capture process without splitting, driven jointly by the magnetic field and the induced electric field. The captured species is selectively determined by the sign of $\vec{E}\cdot\vec{B}$ according to the charge sign, and protons and electrons are accelerated in opposite directions and captured differently.

However, the Wald solution is essentially a test-field approximation and ignores the backreaction of the magnetic field on the spacetime geometry. To overcome the shortcomings of the Wald test-field approximation, exact solutions such as Kerr--Melvin were constructed through the Harrison transformation\cite{19}, describing a Kerr black hole immersed in a Bonnor--Melvin magnetic universe\cite{22,23}, including the backreaction of the magnetic field on the spacetime geometry. However, Kerr--Melvin-type black holes also have some obvious drawbacks. Their Weyl tensor is algebraic type I\cite{24}, rather than the more convenient type D for physical analysis. The ergosphere extends to infinity\cite{25}, making such solutions difficult to use as global magnetized black hole models. Therefore, although Kerr--Melvin black holes are mathematically important, they still have limitations in describing magnetized black holes in realistic astrophysical environments.

To overcome the respective limitations of the Wald solution and Kerr--Melvin-type models, Podolsky and Ovcharenko recently proposed a new class of exact solutions---Kerr--Bertotti--Robinson black holes\cite{1}, which describe a Kerr black hole immersed in an asymptotically uniform external magnetic field along the rotation axis\cite{2,3}. The solution is characterized by three physical parameters: black hole mass $M$, rotation parameter $a$, and external magnetic field strength $B$. It satisfies the full Einstein--Maxwell equations and therefore contains no test-field approximation. When $B\to0$ it directly reduces to the standard Kerr metric in Boyer--Lindquist form; when $M\to0$ it reduces to a Bertotti--Robinson universe filled with a uniform Maxwell field and with $AdS_2\times S^2$ product geometry; when $a\to0$ it reduces to the Schwarzschild--Bertotti--Robinson solution. Its electromagnetic field structure is particularly noteworthy. Unlike the sourced aligned field of the Kerr--Newman solution and the algebraic type~I of the Kerr--Melvin solution, the Kerr--Bertotti--Robinson black hole carries a non-aligned and non-null Maxwell field. In the Newman--Penrose formalism, its non-aligned parts $\Phi_0,\Phi_2$ vanish strictly at the event horizon and on the symmetry axis ($\theta=0,\pi$), while the aligned part $\Phi_1$ is described by a cubic polynomial in $r$, $B_0+B_1r+B_2r^2+B_3r^3$, whose coefficients $B_i$ are functions of $\theta$ only. In addition to the improvement in algebraic structure, the Kerr--Bertotti--Robinson black hole also possesses a series of favorable properties that make it suitable as a global magnetized black hole model: the Weyl tensor remains algebraic type~D, the magnetic field is asymptotically uniform at infinity rather than divergent, the ergosphere is bounded (unlike Kerr--Melvin, which extends to infinity), and both symmetry axes can be regularized simultaneously by a single conical parameter $C$. Therefore, the Kerr--Bertotti--Robinson black hole remedies both the defect of the Wald solution in ignoring the backreaction of the magnetic field on spacetime geometry and the unsuitability of the Kerr--Melvin solution as a global model, providing a more appropriate background spacetime than either for studying magnetized black holes in realistic astrophysical environments.

Although this class of black holes has attracted much attention in many aspects since its proposal\cite{4,26,27,28,30,32,33,34,35,38,39,40,41,43,46,47,49,51,52,55,56,57,58,59,60,61,62,63,29}, its electrodynamic processes, namely the acceleration of electrons and protons by the induced electric field, the selective capture of charged particles in polar and equatorial regions by the black hole, and the resulting energy and angular momentum budgets, have not yet been systematically studied. This study has clear motivation. On the one hand, the region of the black hole magnetosphere where $\vec{E}\cdot\vec{B}\neq 0$ directly determines the acceleration direction and capture geometry of charged particles, and thus whether energy extraction can occur. On the other hand, the electromagnetic four-potential of the Kerr-Bertotti-Robinson black hole is completely different in structure from that of the Wald solution. In particular, since $\Phi_0=\Phi_2$ vanish on the horizon, $\vec{E}\cdot\vec{B}$ at the horizon is completely determined by the aligned part $\Phi_1$, and its $\vec{E}\cdot\vec{B}=0$ critical angle will vary with the dimensionless  magnetic field strength $BM$, rather than taking a single fixed value independent of the magnetic field as in the Kerr black hole in the Wald solution. Furthermore, the two technical premises of the Wald solution no longer automatically hold in the Kerr--Bertotti--Robinson spacetime, which constitutes two points that need to be rehandled in this paper. First, in the Wald solution $A_\phi\simeq \frac{1}{2}B r^2\sin^2\theta$, the magnetic field lines are approximately cylindrical, so that ``magnetic field lines passing through the horizon'' can be directly written as $r\sin\theta\le r_+$. Whereas the Kerr--Bertotti--Robinson spacetime is asymptotically Bertotti--Robinson, and $A_\phi$ saturates at large distances instead of diverging as $r^2$, so the magnetic field lines clearly deviate from cylinders, and the capture criterion must be replaced by the gauge-invariant magnetic flux $A_\phi$ itself. Second, the zero interface of $\vec{E}\cdot\vec{B}=0$ is only an angle $\theta_c$ defined on the horizon. It does not extend as a cone. The interface contracts with radius and may even disappear completely, so the acceleration direction of particles must be determined point by point. In view of this, in this paper we will study the electromagnetic field structure of this black hole and the electrodynamic processes of charged particles. We expect to reveal features different from those of the Kerr black hole in the Wald solution, thereby providing new insights into the properties of this class of magnetized black holes.

The rest of this paper is organized as follows: in Sec.~2, we introduce the electromagnetic field structure of the Kerr--Bertotti--Robinson black hole; in Sec.~3, we analyze the energy and angular momentum of particles, use the magnetic flux criterion to determine the ranges of protons and electrons captured by the black hole, and analyze the energy and angular momentum transferred by the black hole; in Sec.~4, we give conclusions.

\section{Electromagnetic field structure of the Kerr--Bertotti--Robinson black hole}
In Boyer--Lindquist coordinates, the Kerr--Bertotti--Robinson metric is\cite{1}
\begin{equation}
\begin{split}
ds^{2}
&= \frac{1}{\Omega^{2}}\left\{
-\frac{Q-P a^{2}\sin^{2}\theta}{\rho^{2}}\,dt^{2}
+\frac{2a\sin^{2}\theta\left[Q-P(r^{2}+a^{2})\right]}{\rho^{2}}\,dt\,d\phi
\right.\\
&\qquad\qquad
\left.
+\frac{\rho^{2}}{Q}\,dr^{2}
+\frac{\rho^{2}}{P}\,d\theta^{2}
+\frac{\sin^{2}\theta\left[P(r^{2}+a^{2})^{2}-Q a^{2}\sin^{2}\theta\right]}{\rho^{2}}\,d\phi^{2}
\right\}.
\end{split}
\end{equation}
The metric functions are
\begin{equation}
\begin{aligned}
\rho^{2}&=r^{2}+a^{2}\cos^{2}\theta,\quad
P=1+B^{2}\Bigl(M^2\frac{I_{2}}{I_{1}^{2}}-a^{2}\Bigr)\cos^{2}\theta,\quad
Q=(1+B^{2}r^{2})\Delta,\\
\Omega^{2}&=(1+B^{2}r^{2})-B^{2}\Delta\cos^{2}\theta,\quad
\Delta=\Bigl(1-B^{2}M^2\frac{I_{2}}{I_{1}^{2}}\Bigr)r^{2}-2M\frac{I_{2}}{I_{1}}r+a^{2},\\
I_{1}&=1-\tfrac{1}{2}B^{2}a^{2},\quad I_{2}=1-B^{2}a^{2}.
\end{aligned}
\end{equation}
Here, $B$, $a$, and $M$ denote the magnetic field, spin, and mass of the black hole, respectively. The black hole horizon is located at $Q=0$. Solving this condition gives the event horizon $r_+$ and the Cauchy horizon $r_-$
\begin{equation}
r_\pm=\frac{MI_{2}\pm\sqrt{M^2I_{2}-a^{2}I_{1}^{2}}}{I_{1}^{2}-M^2B^{2}I_{2}}I_{1}.
\end{equation}
The Newman--Penrose electromagnetic scalars of the black hole are\cite{1}
\begin{equation}
\Phi_0 = \Phi_2 = B
\frac{1}{2\Omega}
\frac{\sqrt{PQ}\sin\theta}{r + \mathrm{i}a\cos\theta},\quad\Phi_1 = B
\frac{B_0 + B_1 r + B_2 r^2 + B_3 r^3}
{2\Omega I_1^3 (r + \mathrm{i}a\cos\theta)^2},\label{4}
\end{equation}
where
\begin{equation}
\begin{array}{l}
B_0 = a I_1^2 \cos\theta
\left[
M I_2 \cos\theta
- \mathrm{i} a I_1 (1 - B^2 a^2 \cos^2\theta)
\right],\\[4pt]
B_1 = a I_1
\left[
-I_1^2 (1 + \cos^2\theta)
+ B^2 \cos^2\theta
\left(
M^2 I_2 + 2a^2 I_1^2
- \mathrm{i} 3 a M I_1 I_2 \cos\theta
\right)
\right],\\[4pt]
B_2 = I_1
\left[
-3 B^2 a M I_1 I_2 \cos^2\theta
- \mathrm{i} D B^2 a^2
+ \mathrm{i} \cos\theta
\left(
I_1^2
- B^2 M^2 I_2 (1 - 2 I_2 \cos^2\theta)
\right)
\right],\\[4pt]
B_3 = -B^2
\left[
a I_1
\left(
I_1^2 \sin^2\theta
+ B^2 M^2 I_2 \cos^2\theta
\right)
- \mathrm{i} D M I_2
\right],\\[4pt]
D = \cos\theta
\left[
I_1^2 (2 - \cos^2\theta)
+ B^2 M^2 I_2 \cos^2\theta
\right].
\end{array}
\end{equation}
Here, the dual rotation parameter is taken to be $0$. Since $M\neq0$, both electric and magnetic fields are present.

The real part of the electromagnetic four-potential of the black hole is\cite{1}
\begin{equation}
A_t = \frac{a (r \Omega_{,r} + \Omega_{,\theta} \cot\theta)}{B (r^2 + a^2 \cos^2\theta)},\quad
A_\phi = -\frac{r(\Omega_{,r}(a^2 + r^2) - \Omega r + r)}{B(r^2 + a^2 \cos^2\theta)} - \frac{a^2 \cos^2\theta(1 - \Omega) + a^2 \Omega_{,\theta} \sin\theta \cos\theta}{B(r^2 + a^2 \cos^2\theta)}.
\end{equation}
Here, $\Omega_{,r}$ and $\Omega_{,\theta}$ denote partial derivatives of $\Omega$ with respect to $r$ and $\theta$, respectively.

\subsection{Review of the electromagnetic invariant $\vec{E}\cdot\vec{B}$}
In this section, starting from the real electromagnetic tensor, we give the exact relation between the invariant $\vec{E}\cdot\vec{B}$ and the Newman--Penrose scalars $\Phi_i$, and clarify the sign convention. Let the electromagnetic tensor be $F_{\mu\nu}=\partial_\mu A_\nu-\partial_\nu A_\mu$, and define its Hodge dual as
\begin{equation}
{}^\star F^{\mu\nu}=\frac{1}{2}\,\epsilon^{\mu\nu\rho\sigma}F_{\rho\sigma},
\quad \epsilon_{0123}=\sqrt{-g},
\end{equation}
which satisfies ${}^\star({}^\star F)=-F$ under the $(-,+,+,+)$ signature. We further construct the complex self-dual tensor
\begin{equation}
\mathcal{F}_{\mu\nu}=F_{\mu\nu}+\mathrm{i}\,{}^\star F_{\mu\nu},
\quad {}^\star\mathcal{F}_{\mu\nu}=-\mathrm{i}\,\mathcal{F}_{\mu\nu}.
\label{selfdual}
\end{equation}
Contracting it with itself and using ${}^\star F_{\mu\nu}{}^\star F^{\mu\nu}=-F_{\mu\nu}F^{\mu\nu}$ and
$F_{\mu\nu}{}^\star F^{\mu\nu}={}^\star F_{\mu\nu}F^{\mu\nu}$, we obtain
\begin{equation}
\mathcal{F}_{\mu\nu}\mathcal{F}^{\mu\nu}
=2\,F_{\mu\nu}F^{\mu\nu}+2\,\mathrm{i}\,F_{\mu\nu}{}^\star F^{\mu\nu}.
\label{FF}
\end{equation}
In any orthonormal local frame, the two real invariants are
\begin{equation}
F_{\mu\nu}F^{\mu\nu}=2\,(\vec{B}^{2}-\vec{E}^{2}),\quad
F_{\mu\nu}{}^\star F^{\mu\nu}=4\,\vec{E}\cdot\vec{B}.
\label{twoinv}
\end{equation}
Substituting into Eq.~\eqref{FF} gives
\begin{equation}
\mathcal{F}_{\mu\nu}\mathcal{F}^{\mu\nu}
=4\,(\vec{B}^{2}-\vec{E}^{2})+8\,\mathrm{i}\,\vec{E}\cdot\vec{B}.
\label{FF2}
\end{equation}
On the other hand, Ref.~\cite{1} gives the relation between the complex self-dual tensor and the Newman--Penrose scalars:
\begin{equation}
\frac{1}{16}\,\mathcal{F}_{\mu\nu}\mathcal{F}^{\mu\nu}=\Phi_0\Phi_2-\Phi_1^2 .
\label{NPrel}
\end{equation}
Comparing Eq.~\eqref{FF2} with Eq.~\eqref{NPrel} and taking the real and imaginary parts separately, we obtain
\begin{equation}
\vec{E}\cdot\vec{B}=2\,\operatorname{Im}\!\left(\Phi_0\Phi_2-\Phi_1^2\right).
\label{6}
\end{equation}
Equation~\eqref{6} is the electromagnetic invariant used in this paper. On the event horizon $Q=0$, from Eq.~\eqref{4} we have $\Phi_0=\Phi_2=0$, so Eq.~\eqref{6} becomes
\begin{equation}
(\vec{E}\cdot\vec{B})_H=-\,2\,\operatorname{Im}\Phi_1^2 ,
\label{EBH}
\end{equation}
i.e., the electromagnetic invariant at the horizon is completely determined by the aligned part $\Phi_1$.

Following Refs.~\cite{5,6}, we solve at the event horizon for the critical angle $\theta_c$ satisfying $\vec{E}\cdot\vec{B}=0$. This angle is the boundary between the regions where the scalar product is positive and negative. For spin parameter $a/M=0.7$, the value of $\theta_c$ for $\vec{E}\cdot\vec{B}=0$ depends on the dimensionless magnetic field $BM$, rather than being the fixed value $56.12^\circ$\cite{5} of the Kerr black hole in the Wald solution. In Fig.~\ref{fig:1} we plot the variation of different critical angles $\theta_c$ with the magnetic field $BM$.

It is necessary to state the sign convention explicitly: if one takes the self-dual tensor as
$\mathcal{F}'_{\mu\nu}=F_{\mu\nu}-\mathrm{i}\,{}^\star F_{\mu\nu}$ (or takes the opposite Levi-Civita orientation), the imaginary part in Eq.~\eqref{FF2} changes sign, yielding
$\vec{E}\cdot\vec{B}=-2\,\operatorname{Im}(\Phi_0\Phi_2-\Phi_1^2)$. Under the two conventions, the zeros of
$\operatorname{Im}(\Phi_0\Phi_2-\Phi_1^2)=0$ are exactly the same, so all conclusions in this paper concerning the critical angle
$\theta_c$ are independent of the sign convention. However, when determining the particle acceleration direction, one must use the same convention as for $\vec{E}\cdot\vec{B}$ in the ZAMO frame below. In this paper we uniformly adopt Eq.~\eqref{6}.

\subsection{Electric and magnetic fields in the ZAMO frame}
To determine the acceleration direction of charged particles, it is necessary to give the components of $\vec{E}$ and $\vec{B}$ in a local inertial frame. We adopt the locally non-rotating frame (ZAMO). Writing the metric as
\begin{equation}
ds^2=g_{tt}dt^2+2g_{t\phi}dtd\phi+g_{rr}dr^2+g_{\theta\theta}d\theta^2+g_{\phi\phi}d\phi^2 ,
\end{equation}
the orthonormal ZAMO tetrad is chosen as
\begin{equation}
e_{\hat 0}=\frac{1}{N}\left(\partial_t+\omega\,\partial_\phi\right),\quad
e_{\hat 1}=\frac{1}{\sqrt{g_{rr}}}\partial_r,\quad
e_{\hat 2}=\frac{1}{\sqrt{g_{\theta\theta}}}\partial_\theta,\quad
e_{\hat 3}=\frac{1}{\sqrt{g_{\phi\phi}}}\partial_\phi,
\label{tetrad}
\end{equation}
where
\begin{equation}
\omega=-\frac{g_{t\phi}}{g_{\phi\phi}},\quad
N=\sqrt{g_{\phi\phi}\omega^2-g_{tt}}
=\sqrt{\frac{g_{t\phi}^2}{g_{\phi\phi}}-g_{tt}}
\label{lapse}
\end{equation}
are the ZAMO angular velocity and lapse function, respectively, satisfying $e_{\hat 0}\cdot e_{\hat 0}=-1$. Because of axisymmetry and stationarity, $A_\mu=(A_t,0,0,A_\phi)$, and the nonzero components of $F_{\mu\nu}$ are
\begin{equation}
F_{rt}=\partial_r A_t,\quad F_{\theta t}=\partial_\theta A_t,\quad
F_{r\phi}=\partial_r A_\phi,\quad F_{\theta\phi}=\partial_\theta A_\phi .
\end{equation}
From the tetrad components $F_{\hat a\hat b}=F_{\mu\nu}e_{\hat a}^{\ \mu}e_{\hat b}^{\ \nu}$,
and defining the local electromagnetic field by $E_{\hat i}=F_{\hat i\hat 0}$ and $F_{\hat i\hat j}=\epsilon_{\hat i\hat j\hat k}B_{\hat k}$, we obtain the electric field components
\begin{equation}
E_{\hat r}=\frac{\partial_r A_t+\omega\,\partial_r A_\phi}{N\sqrt{g_{rr}}},\quad
E_{\hat\theta}=\frac{\partial_\theta A_t+\omega\,\partial_\theta A_\phi}{N\sqrt{g_{\theta\theta}}},\quad
E_{\hat\phi}=0,
\label{ZAMOE}
\end{equation}
and the magnetic field components
\begin{equation}
B_{\hat r}=\frac{\partial_\theta A_\phi}{\sqrt{g_{\theta\theta}g_{\phi\phi}}},\quad
B_{\hat\theta}=-\frac{\partial_r A_\phi}{\sqrt{g_{rr}g_{\phi\phi}}},\quad
B_{\hat\phi}=0.
\label{ZAMOB}
\end{equation}
Thus the electromagnetic invariant is
\begin{equation}
\vec{E}\cdot\vec{B}=E_{\hat r}B_{\hat r}+E_{\hat\theta}B_{\hat\theta}
=\frac{\partial_r A_t\,\partial_\theta A_\phi-\partial_\theta A_t\,\partial_r A_\phi}
{N\sqrt{g_{rr}g_{\theta\theta}g_{\phi\phi}}}.
\label{EBcompact}
\end{equation}
It is worth noting that the two terms containing $\omega$ in Eq.~\eqref{ZAMOE} cancel exactly in Eq.~\eqref{EBcompact}.
Therefore, $\vec{E}\cdot\vec{B}$ is independent of frame dragging, which is exactly the property it should have as a scalar invariant.
We have checked Eq.~\eqref{EBcompact} against Eq.~\eqref{6} pointwise numerically; the two agree completely at all sampled points,
thereby confirming the $+2$ sign in Eq.~\eqref{6}.

\subsection{Criterion for the acceleration direction of particles}
\begin{figure}[!h]
  \centering
    \includegraphics[width=0.45\linewidth]{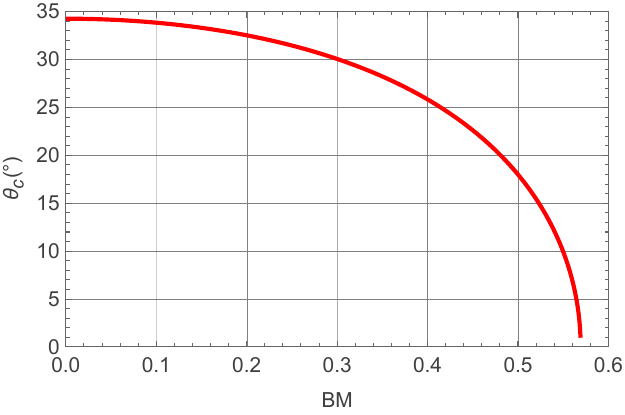}
 \caption{At the event horizon, the variation of the critical angle $\theta_c$ with the dimensionless magnetic field $BM$. The dimensionless spin is taken as $a/M=0.7$ following Ref.~\cite{5}. In this case the extremal black hole is located at $BM=1.3$, i.e., all magnetic fields satisfy the condition for horizon existence, ensuring that no naked singularity appears.}
\label{fig:1}
\end{figure}

Figure~\ref{fig:1} only gives the dividing angle $\theta_c$ between the polar and equatorial regions. To determine toward which end a charged particle is actually pushed, one must also take into account the sign of the particle charge. In the magnetically dominated case ($B^2>E^2$), charged particles are confined to move along magnetic field lines, and their acceleration comes entirely from the electric field component parallel to the magnetic field lines. The parallel force is along the magnetic field line direction, and the sign of its radial component is
\begin{equation}
\operatorname{sgn}\!\left(F_\parallel^{\hat r}\right)
=\operatorname{sgn}\!\left[\,q_i\,(\vec{E}\cdot\vec{B})\,B_{\hat r}\right].
\label{criterion}
\end{equation}
Introducing
\begin{equation}
S\equiv(\vec{E}\cdot\vec{B})\,B_{\hat r} ,
\label{Sdef}
\end{equation}
when $S>0$, protons ($q=+e$) are pushed toward infinity (outward) and electrons ($q=-e$) toward the horizon (inward); when $S<0$, their roles are interchanged. Note that $q_i(\vec{E}\cdot\vec{B})B_{\hat r}$ in Eq.~\eqref{criterion} is invariant under the overall gauge transformation $A_\mu\to-A_\mu$, so the physical conclusions obtained are independent of the gauge choice. From Eq.~\eqref{4} and Eq.~\eqref{6}, near the horizon we have
\begin{equation}
\begin{aligned}
\text{polar region }(0<\theta<\theta_c):\quad
  &\vec{E}\cdot\vec{B}<0,\ B_{\hat r}<0\ \Rightarrow\ S>0;\\
\text{equatorial region }(\theta_c\le\theta\le\pi/2):\quad
  &\vec{E}\cdot\vec{B}>0,\ B_{\hat r}<0\ \Rightarrow\ S<0 .
\end{aligned}
\label{signpolar}
\end{equation}
Here $B_{\hat r}<0$ is a substantial difference from the Wald solution: in the Wald solution
$B_{\hat r}^H=2\cos^2\theta\,BMr_+(r_+^2-a^2)/\Sigma_H^2>0$\cite{5}, whereas
for the Kerr--Bertotti--Robinson black hole, $A_\phi$ has the opposite sign to that in the Wald solution and decreases monotonically with $\theta$,
so $B_{\hat r}=\partial_\theta A_\phi/\sqrt{g_{\theta\theta}g_{\phi\phi}}<0$.
From Eq.~\eqref{signpolar} and Eq.~\eqref{criterion}, we obtain: in the polar region near the horizon, protons are accelerated outward and escape, while electrons are accelerated inward and fall into the black hole; in the equatorial region, their roles are interchanged, with electrons escaping outward and protons falling inward. The southern hemisphere
($\pi-\theta_c<\theta<\pi$) gives the same conclusion by symmetry: under $\theta\to\pi-\theta$,
$\Phi_1\to\bar\Phi_1$, i.e., $E_{\hat r}$ is unchanged, while $B_{\hat r}$ and
$\vec{E}\cdot\vec{B}$ change sign simultaneously, so $S$ does not change sign. Reversing the sign of $B$ (or $a$) interchanges the roles of protons and electrons globally, but does not change the conclusions below regarding the energy budget.

According to Fig.~\ref{fig:1}, when $BM\to0$, $\theta_c\approx34.26^\circ$. As the magnetic field increases, the polar region shrinks and the equatorial region expands. When $BM\approx0.569$, $\theta_c\approx0$, indicating that the polar region at the horizon almost disappears.

However, it must be emphasized that $\theta_c$ is only an angle defined on $r=r_+$, and the ``polar region'' and ``equatorial region'' divided by it do not constitute cones. This point is not obvious in the Wald solution: the Wald solution is asymptotically flat, and the zero interface of $\vec{E}\cdot\vec{B}=0$ is approximately the cone $\theta=\theta_c$, so a single angle can be used to divide the entire space. The Kerr--Bertotti--Robinson spacetime is asymptotically Bertotti--Robinson, with $\Omega\simeq Br\sqrt{1-k\cos^2\theta}$ growing linearly with $r$ ($k\equiv1-B^2M^2I_2/I_1^2$), and the zero interface contracts monotonically with radius. In Table~\ref{tab:2} we give the variation of $\theta_c$ with radius for $BM=0.3$: it decreases monotonically from $30.06^\circ$ at the horizon to about $9^\circ$ at $r\simeq2.15\,r_+$, and shrinks to the axis at $r=2.2517\,r_+$, after which it disappears completely. At that point $\vec{E}\cdot\vec{B}$ no longer changes sign over the entire $(r,\theta)$ plane, and the polar region no longer exists. Therefore, the acceleration direction of particles must be determined point by point using Eq.~\eqref{criterion}, and one cannot simply use the single angle $\theta_c$ on the horizon. This effect will play a decisive role in the energy budget in Sec.~3: it is precisely because the polar region collapses first in the outer region that negative-energy protons can be captured.

\begin{table}[htbp]
    \centering
    \caption{Variation of the critical angle $\theta_c$ for $\vec{E}\cdot\vec{B}=0$ with radius, for $BM=0.3,\ a/M=0.7$ (here $r_+=1.8584$). The zero interface is not a cone: $\theta_c$ decreases monotonically with radius and closes continuously to zero at
    $r\simeq2.252\,r_+$, after which the polar region no longer exists.}
    \label{tab:2}
    \begin{tabular}{cccccccccc}
        \toprule
        $r/r_+$ & 1.00 & 1.10 & 1.30 & 1.50 & 1.80 & 2.00 & 2.15 & 2.25 & $\gtrsim2.252$ \\
        \midrule
        $\theta_c$ ($^\circ$) & 30.06 & 29.31 & 27.23 & 24.50 & 19.14 & 14.29 & 9.07 & 1.17 & no zero point \\
        \bottomrule
    \end{tabular}
\end{table}

\begin{figure}[H]
  \centering
  \begin{subfigure}{0.45\textwidth}
    \centering
    \includegraphics[width=\linewidth]{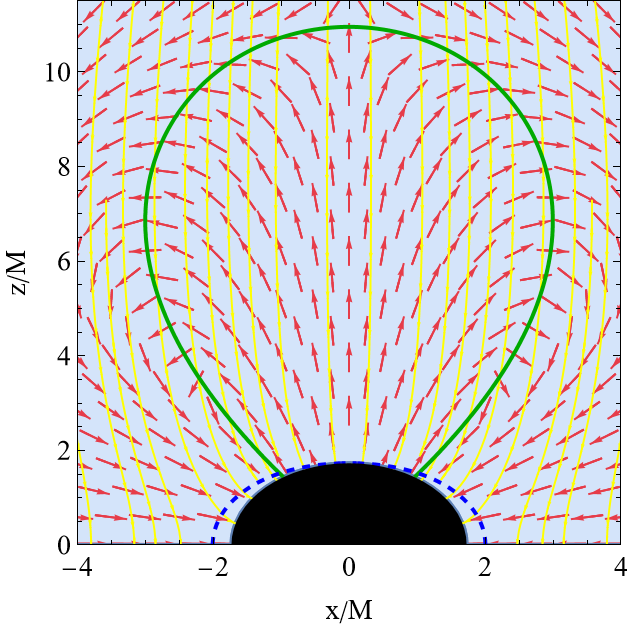}
    \caption{$BM=0.1$}
  \end{subfigure}
  \hfill
  \begin{subfigure}{0.45\textwidth}
    \centering
    \includegraphics[width=\linewidth]{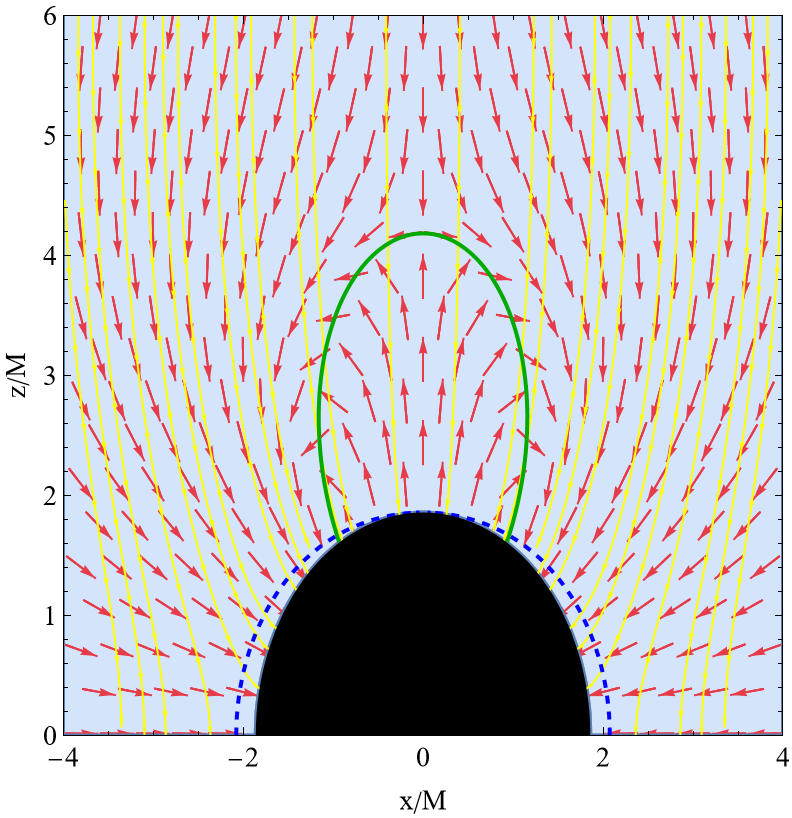}
    \caption{$BM=0.3$}
  \end{subfigure}
  \begin{subfigure}{0.45\textwidth}
    \centering
    \includegraphics[width=\linewidth]{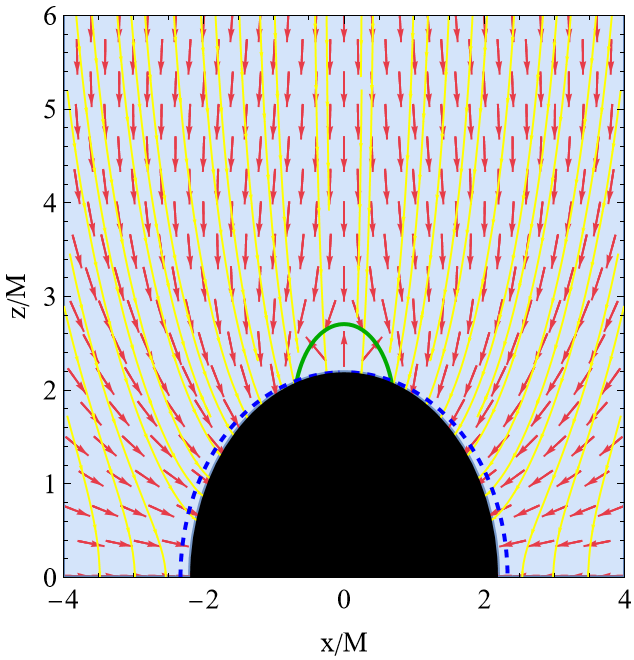}
    \caption{$BM=0.5$}
  \end{subfigure}
  \hfill
  \begin{subfigure}{0.45\textwidth}
    \centering
    \includegraphics[width=\linewidth]{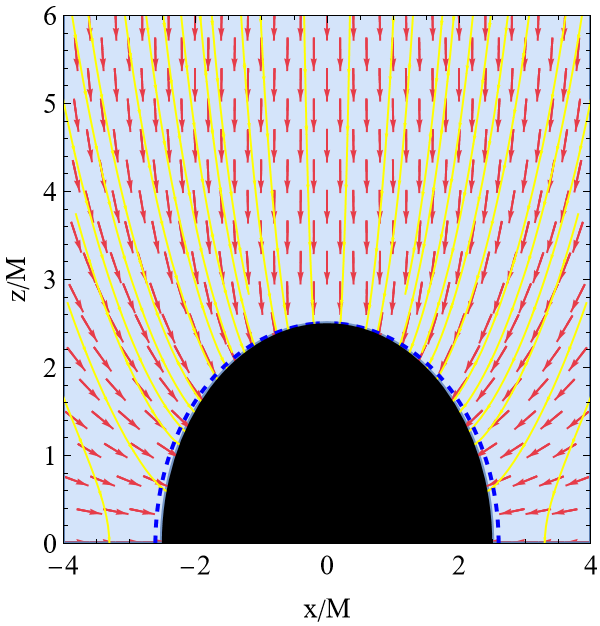}
    \caption{$BM=0.6$}
  \end{subfigure}
\caption{Electromagnetic field structure of the Kerr--Bertotti--Robinson black hole on the meridional plane $(x/M,\,z/M)$, with $a/M=0.7$,
in the northern hemisphere. The southern hemisphere can be obtained by mirroring the northern hemisphere about the equatorial plane, with the only exception being the magnetic field direction: magnetic field lines run from north to south through both hemispheres and are not mirrored. The black disk is the interior of the event horizon; the blue dashed line is the ergosphere boundary; the green solid line is the zero interface of $\vec{E}\cdot\vec{B}=0$; the red arrows are the electric field in the ZAMO frame; the yellow directed streamlines are the magnetic field. As the magnetic field increases, the polar region enclosed by the zero interface shrinks monotonically and completely collapses at $BM\simeq0.569$.}
\label{fig:2}
\end{figure}

To display the above electromagnetic field structure more intuitively, we give in Fig.~\ref{fig:2} the image on the meridional plane $(x/M,\,z/M)$,
where
\begin{equation}
x=r\sin\theta,\quad z=r\cos\theta ,
\end{equation}
and the direction is in the northern hemisphere ($z\ge0$). In the figure, the black disk is the interior of the event horizon ($r<r_+$), the blue dashed line is the ergosphere boundary ($g_{tt}=0$),
the green solid line is the zero interface of $\vec{E}\cdot\vec{B}=0$, the red discrete arrows give the electric field $\vec{E}$ in the ZAMO frame,
and the yellow streamlines with directions give the magnetic field $\vec{B}$.
Three structural features can be read from Fig.~\ref{fig:2}. First, the zero interface is an arch-shaped curve that starts from the horizon, bulges outward, and finally closes on the symmetry axis, rather than being a cone
$\theta=\theta_c$. This is completely consistent with the monotonic contraction of $\theta_c(r)$ given in Table~\ref{tab:2}:
$\theta_c$ takes its maximum on the horizon and decreases with increasing radius until it tends to zero at some radius, where the arch closes on the axis.
Numerically, the closing point is located at $z\simeq10.95$ for $BM=0.1$, $z\simeq4.19$ for $BM=0.3$, and
$z\simeq2.71$ for $BM=0.5$; the widest parts of the arch are at $x\simeq\pm3.00$, $\pm1.16$, and $\pm0.68$, respectively. Second, the relative orientation of the electric and magnetic fields is opposite in the polar and equatorial regions. From Eq.~\eqref{ZAMOB},
$B_{\hat r}=\partial_\theta A_\phi/\sqrt{g_{\theta\theta}g_{\phi\phi}}<0$ holds at the horizon and throughout space
(because $A_\phi$ has the opposite sign to that in the Wald solution and decreases monotonically with $\theta$). Hence, in the northern hemisphere the magnetic field is globally along the $-z$ direction,
i.e., antiparallel to the black hole spin, consistent with the magnetic field lines running downward in the figure and with the far-field arrows pointing downward.
In the polar region (inside the arch), $E_{\hat r}>0$, so the electric field points upward along $+z$, opposite to the magnetic field,
and therefore $\vec{E}\cdot\vec{B}<0$; in the equatorial region (outside the arch), $E_{\hat r}<0$, so the electric field and magnetic field are in the same direction,
and therefore $\vec{E}\cdot\vec{B}>0$. Combining this with $B_{\hat r}<0$, Eq.~\eqref{criterion} gives
$S>0$ in the polar region (protons outward, electrons inward) and $S<0$ in the equatorial region (electrons outward, protons inward),
in complete agreement with the previous analysis. Third, and most importantly, the polar region collapses monotonically as the magnetic field increases, until it disappears completely.
When $BM=0.1$, the critical angle at the horizon is $33.85^\circ$, and the arch is wide; when $BM=0.3$, it decreases to $30.06^\circ$;
when $BM=0.5$, it is only $18.00^\circ$, and the arch has degenerated into a small region clinging to the horizon.
When $BM=0.6$, no zero interface is visible in the figure: at this time $\vec{E}\cdot\vec{B}>0$ holds over the entire
$(r,\theta)$ plane, and the polar region no longer exists. This collapse occurs at
\begin{equation}
BM\simeq0.569\quad (a/M=0.7),
\end{equation}
consistent with the location where $\theta_c\to0$ in Fig.~\ref{fig:1}.

The third point above will have a decisive influence on the energy budget in Sec.~3: once the polar region completely disappears,
the particles pushed toward the horizon by the parallel electric field will all be protons, and protons in this model carry negative energy and negative angular momentum. Therefore, under a sufficiently strong magnetic field, the net energy and net angular momentum captured by the black hole can both be negative,
and energy extraction can be realized without invoking any special anisotropic density distribution.

\section{Energy and angular momentum of particles}
The conserved energy and angular momentum of a test particle with mass \(m_i\) and charge \(q_i\) can be expressed as
\begin{equation}
E_i  = -m_i u_t - q_i A_t,\quad L_i  = m_i u_\phi + q_i A_\phi,
\end{equation}
where \(u_\alpha\) is the covariant four-velocity, and \(i = \mathrm{proton},\ \mathrm{electron}\) denotes a proton or an electron. We consider particles initially at rest, i.e., particles located outside the ergosphere. Taking $e$ as the elementary charge, for an electron $q_i=-e$, and for a proton $q_i=e$. Initial rest means $u^r=u^\theta=u^\phi=0$, so from the normalization condition we obtain $u^t = 1/\sqrt{(-g_{tt})}$. Since
\begin{equation}
u_t = g_{tt}u^t + g_{t\phi}u^\phi = g_{tt}u^t,\quad u_\phi = g_{\phi t}u^t + g_{\phi\phi}u^\phi = g_{\phi t}u^t,
\end{equation}
the energy and angular momentum at the initial position can be expressed as
\begin{equation}
\begin{aligned} E_i &= m_i\frac{\sqrt{Q-Pa^2\sin^2\theta}}{\Omega\rho} \pm \left( -\frac{e aB}{\Omega\rho^2} \left[ r^2+\left(a^2-M\frac{I_2}{I_1}r\right)\cos^2\theta \right] \right), \\[2mm] L_i &= m_i\frac{a\sin^2\theta\left[Q-P(r^2+a^2)\right]} {\Omega\rho\sqrt{Q-Pa^2\sin^2\theta}} \\ &\quad \pm \Bigg( -\frac{e}{B\rho^2} \Bigg\{ \frac{r(a^2+r^2)B^2}{\Omega} \left[ r-\left(\left(1-B^2M^2\frac{I_2}{I_1^2}\right)r-M\frac{I_2}{I_1}\right)\cos^2\theta \right] \\ &\quad +(1-\Omega)\rho^2 +\frac{a^2B^2\Delta}{\Omega}\sin^2\theta\cos^2\theta \Bigg\} \Bigg). \end{aligned}\label{10}
\end{equation}
Here, $+$ represents a proton and $-$ represents an electron. Since in realistic astrophysical environments the electromagnetic term is dominant, the mass term can be neglected, so in Eq.~\eqref{10} only the terms on the right of $\pm$ need be retained, and the terms on the left can be ignored. This is consistent with the treatment in Ref.~\cite{5}. Then
\begin{equation}
E_p=-eA_t,\quad L_p=eA_\phi;\qquad E_e=+eA_t,\quad L_e=-eA_\phi.
\label{Ep}
\end{equation}
Under our chosen $a>0,\ B>0$, we have $A_t>0$, so protons carry negative energy and electrons carry positive energy; moreover, $E_e=-E_p$ and $L_e=-L_p$.

The kinetic energy of a particle crossing the event horizon can be expressed as
\begin{equation}
K_i = E_i - \Omega_H L_i ,  \quad\Omega_H = -\left. \frac{g_{t\phi}}{g_{\phi\phi}} \right|_H  =\frac{a}{r_+^2 + a^2}.\label{11}
\end{equation}
Here, $\Omega_H$ is the angular velocity of the event horizon. From Eq.~\eqref{Ep} we immediately obtain $K_e=-K_p$, so $K_p\ge0$ and $K_e\ge0$ are mutually complementary sets.

\subsection{Capture domain: magnetic flux criterion}
Particles move along magnetic field lines (contours of constant $A_\phi$), so for a particle to be captured by the black hole, the magnetic field line on which it lies must pass through the horizon. In the Wald solution $A_\phi\simeq\frac12Br^2\sin^2\theta$, magnetic field lines are approximately cylindrical surfaces $r\sin\theta=\text{constant}$, and ``magnetic field lines passing through the horizon'' is equivalent to the simple geometric condition $r\sin\theta\le r_+$. Reference \cite{5} accordingly takes the initial position $r_i\sin\theta_i=r_+$. However, in the Kerr--Bertotti--Robinson spacetime this criterion fails. Since $\Omega\simeq Br\sqrt{1-k\cos^2\theta}$ grows linearly with $r$, the leading terms in the expression for $A_\phi$ that grow with $r$ cancel each other, and $A_\phi$ does not diverge as $r^2$ as $r\to\infty$, but instead saturates to a constant depending only on $\theta$. Table~\ref{tab:3} gives the saturated values of $A_\phi$ at $r=10^6$ for $BM=0.3$. At this time magnetic field lines clearly deviate from cylinders, and $r\sin\theta$ is no longer a label of magnetic field lines.

\begin{table}[htbp]
    \centering
    \caption{Saturated values $A_\phi^\infty(\theta)$ of $A_\phi$ at $r=10^6$ for $BM=0.3,\ a/M=0.7$ (compared with the horizon range $[A^{H}_{\min},A^{H}_{\max}]=[-0.5503,-0.0743]$).}
    \label{tab:3}
    \begin{tabular}{cccccccc}
        \toprule
        $\theta$ ($^\circ$) & 0.1 & 1 & 5 & 10 & 20 & 30 & 90 \\
        \midrule
        $A_\phi^\infty$ & $-0.0744$ & $-0.0803$ & $-0.2166$ & $-0.5666$ & $-1.3858$ & $-2.0322$ & $-3.3333$ \\
        \bottomrule
    \end{tabular}
\end{table}

The correct criterion should be given directly by the gauge-invariant magnetic flux. Let the range of $A_\phi$ on the horizon be
\begin{equation}
A_\phi\in\bigl[A^{H}_{\min},\,A^{H}_{\max}\bigr],\qquad
A^{H}_{\min/\max}=\min_{\theta}/\max_{\theta}\,A_\phi(r_+,\theta),\label{band}
\end{equation}
then a magnetic field line passes through the horizon if and only if its $A_\phi$ value lies in the interval~\eqref{band}. Since $A_\phi$ decreases monotonically along a ray, this condition gives the outer boundary $r_{\rm out}(\theta)$ on each ray, determined by $A_\phi(r_{\rm out},\theta)=A^{H}_{\min}$. We compare the true $r_{\rm out}(\theta)$ with the Wald-type $r_+/\sin\theta$ in Table~\ref{tab:4}. The difference is significant: for example, at $\theta=30^\circ$ the true outer boundary is $1.40$ times $r_+/\sin\theta$, and at $\theta=15^\circ$ it reaches $2.79$ times; while for $\theta<\theta^\ast$, $A_\phi^\infty(\theta)>A^{H}_{\min}$, the magnetic field line extends to infinity while still passing through the horizon, $r_{\rm out}\to\infty$, and there is no outer boundary at all. Here $\theta^\ast$ is defined by $A_\phi^\infty(\theta^\ast)=A^{H}_{\min}$; it increases rapidly with the magnetic field. Its numerical variation and relation to Meissner-type behavior\cite{1,43,36} will be discussed below together with Fig.~\ref{fig:2}.

\begin{table}[htbp]
    \centering
    \caption{Comparison of the true outer boundary $r_{\rm out}(\theta)$ with the Wald-type $r_+/\sin\theta$ for $BM=0.3,\ a/M=0.7$ (for $BM=0.3$, $\theta^\ast\simeq9.80^\circ$).}
    \label{tab:4}
    \begin{tabular}{ccccccc}
        \toprule
        $\theta$ ($^\circ$) & 1 & 5 & 15 & 30 & 45 & 60 \\
        \midrule
        $r_{\rm out}$ & $\infty$ & $\infty$ & 20.05 & 5.20 & 3.10 & 2.32 \\
        $r_+/\sin\theta$ & 106.5 & 21.3 & 7.18 & 3.72 & 2.63 & 2.15 \\
        $r_{\rm out}\sin\theta/r_+$ & $\infty$ & $\infty$ & 2.79 & 1.40 & 1.18 & 1.08 \\
        \bottomrule
    \end{tabular}
\end{table}

The above criterion can also be equivalently expressed in terms of the polar angle at infinity: since $A_\phi^\infty(\theta)$ is monotonic, condition~\eqref{band} is equivalent to the magnetic field line's polar angle at infinity satisfying $\theta_\infty<\theta^\ast$; here $\theta^\ast$ is the opening angle of the magnetic field line at infinity, not its polar angle at the horizon. According to this criterion, the magnetic field lines in Fig.~\ref{fig:2} are clearly divided into two classes. The first class passes through the horizon. When $\theta_\infty<\theta^\ast$, the $A_\phi$ of the magnetic field line falls within the horizon band, and the magnetic field line passes through the horizon and plunges into it. Taking a magnetic field line with $\theta_\infty\simeq2.0^\circ$ for $BM=0.3$ as an example, its cylindrical radius $R=r\sin\theta$ contracts monotonically from about $3.5\times10^{4}$ at $r=10^6$ to $0.44$ at the horizon, showing a form of being ``attracted'' and converging by the horizon. There is a point that is easily confused and needs special explanation: in the process of falling toward the black hole, magnetic field lines spread significantly toward the equator, and their polar angle at the horizon $\theta_H$ is much larger than their polar angle at infinity $\theta_\infty$. Taking $BM=0.3$ as an example, magnetic field lines with $\theta_\infty\simeq1.26^\circ,\,4.02^\circ,\,6.32^\circ,\,8.14^\circ,\,9.24^\circ,\,9.77^\circ$ fall on the horizon at $\theta_H\simeq8.68^\circ,\,27.91^\circ,\,44.79^\circ,\,60.30^\circ,\,73.26^\circ,\,86.21^\circ$, respectively. Therefore, although the magnetic field lines that can pass through the horizon occupy only a narrow cone with half-opening angle $\theta^\ast\simeq9.80^\circ$ at infinity, when they reach the horizon they cover the entire horizon surface from the polar region to the equator. This is the reason why in Fig.~\ref{fig:2} one can see magnetic field lines with large angles (even close to the equatorial plane) still plunging into the horizon: their $\theta_\infty$ are actually all less than $9.80^\circ$. The second class of magnetic field lines is clearly pushed outward. When $\theta_\infty>\theta^\ast$, the $A_\phi$ of the magnetic field line falls outside the horizon band, cannot pass through the horizon, and can only gradually deflect outward as it falls from afar, bypass the black hole, and finally cross the equatorial plane. Its equatorial crossing radius $r_{\rm eq}$ increases as $A_\phi$ deviates from the horizon band (for $BM=0.3$, taking $\Delta A_\phi\equiv A^{H}_{\min}-A_\phi$ as $0.05,\,0.3,\,0.5,\,1.0$ gives $r_{\rm eq}\simeq2.00,\,2.68,\,3.25,\,4.94$), and except for a few lines immediately adjacent to the horizon band, most have completed their deflection outside the ergosphere ($r_{\rm ergo}(90^\circ)\simeq2.08$). Such magnetic field lines are blocked by the horizon and go around it, which is a direct manifestation of the inability of the magnetic field to enter the horizon.

It should be emphasized that $\theta^\ast$ increases with increasing magnetic field (for $BM=0.1,\,0.3,\,0.6$ it is about $1.07^\circ,\,9.80^\circ,\,36.61^\circ$, respectively), i.e., the bundle of magnetic field lines passing through the horizon becomes wider as the magnetic field increases, rather than contracting toward the symmetry axis. This does not contradict Meissner-type behavior\cite{1,43,36}: the latter measures the magnetic flux passing through the horizon per unit external magnetic field, and we find that $|\Phi_H|/(BM^2)$ decreases monotonically with $BM$ (about $10.68$ at $BM=0.1$ and about $5.16$ at $BM=0.9$), indicating that relative to the external magnetic field strength, the magnetic field is indeed gradually expelled from the horizon. The two describe the geometric aspect of magnetic field line connectivity and the physical aspect of magnetic flux strength, respectively.

\subsection{Two-dimensional capture domain and energy budget}
In summary, a particle must simultaneously satisfy four conditions to be captured by the black hole: (i) located outside the ergosphere, $r>r_{\rm ergo}(\theta)$; (ii) the magnetic field line on which it lies passes through the horizon, $A_\phi\in[A^{H}_{\min},A^{H}_{\max}]$, i.e., $r<r_{\rm out}(\theta)$; (iii) pushed toward the horizon by the parallel electric field, i.e., $q_i S<0$ ($S\equiv(\vec{E}\cdot\vec{B})B_{\hat r}$); (iv) can actually reach the horizon, $K_i\ge0$. Using $E_e=-E_p$, $L_e=-L_p$, and $K_e=-K_p$, the last two conditions can be unified as
\begin{equation}
(\vec{E}\cdot\vec{B})\,B_{\hat r}\,K_p\;\le\;0,
\label{unified}
\end{equation}
and when $(\vec{E}\cdot\vec{B})B_{\hat r}<0$, what is captured is a proton (carrying negative energy); when $(\vec{E}\cdot\vec{B})B_{\hat r}>0$, what is captured is an electron (carrying positive energy). Since $\vec{E}\cdot\vec{B}$ and $B_{\hat r}$ both depend on $(r,\theta)$, condition~\eqref{unified} must be determined point by point on the $(r,\theta)$ plane, and the capture domain $\mathcal D_i$ is a two-dimensional region rather than an angular interval.

The energy and angular momentum transferred by the black hole are
\begin{align}
\mathcal E_i \approx 2\pi C\iint_{\mathcal D_i}
E_i\,n\,\sqrt{-g}\,dr\,d\theta,\quad
\mathcal L_i \approx 2\pi C\iint_{\mathcal D_i}
L_i\,n\,\sqrt{-g}\,dr\,d\theta,
\end{align}
where
\begin{equation}
C=\left[1+B^2\Bigl(M^2\frac{I_2}{I_1^2}-a^2\Bigr)\right]^{-1},\quad\sqrt{-g}=\frac{\rho^2\sin\theta}{\Omega^4}.
\end{equation}
Here, $C$ is the conical parameter\cite{1}, and $\sqrt{-g}$ is the metric determinant. We adopt a spherically symmetric density\cite{6}
\begin{equation}
    n(r)=n_+(r_+/r)^3.
\end{equation}
All integrals below are summed over both the northern and southern hemispheres: under $\theta\to\pi-\theta$, $A_t,A_\phi$ are unchanged while $B_{\hat r}$ and $\vec{E}\cdot\vec{B}$ change sign simultaneously, so $S$ and $K_p$ do not change sign, and the contributions of the two hemispheres are strictly equal. The integrand decays rapidly as $r^{-6}$, so the integral is insensitive to the radial outer boundary and is numerically convergent.

\begin{table}[htbp]
    \centering
    \caption{For spherically symmetric density $n=n_+(r_+/r)^3$ and $a/M=0.7$, the energy and angular momentum obtained by the black hole as functions of the magnetic field. $\Delta\mathcal E>0$ means the black hole mass increases, and $\Delta\mathcal E<0$ means energy extraction. The transition occurs at $BM\approx0.551$; the electron capture domain closes together with the polar region at $BM\simeq0.569$, after which $\mathcal E_e=0$.}
    \label{tab:5}
    \begin{tabular}{ccccc}
        \toprule
        $BM$ & $\mathcal{E}_p/(eBan_+M^3)$ & $\mathcal{E}_e/(eBan_+M^3)$ & $\Delta\mathcal{E}/(eBan_+M^3)$ & $\Delta\mathcal{L}/(eBn_+M^5)$ \\
        \midrule
        0.1   & $-1.31\times10^{-4}$ & $7.668$ & $+7.668$ & $+9.554$ \\
        0.3   & $-6.86\times10^{-3}$ & $1.375$ & $+1.369$ & $+1.121$ \\
        0.5   & $-5.30\times10^{-3}$ & $5.22\times10^{-2}$ & $+4.69\times10^{-2}$ & $+3.02\times10^{-3}$ \\
        0.54   & $-3.74\times10^{-3}$ & $8.50\times10^{-3}$ & $+4.76\times10^{-3}$ & $-2.03\times10^{-2}$ \\
        0.55   & $-3.39\times10^{-3}$ & $3.61\times10^{-3}$ & $+2.15\times10^{-4}$ & $-2.18\times10^{-2}$ \\
        0.56   & $-3.06\times10^{-3}$ & $8.15\times10^{-4}$ & $-2.25\times10^{-3}$ & $-2.19\times10^{-2}$ \\
        0.565   & $-2.90\times10^{-3}$ & $1.70\times10^{-4}$ & $-2.73\times10^{-3}$ & $-2.15\times10^{-2}$ \\
        0.569   & $-2.78\times10^{-3}$ & $3.76\times10^{-7}$ & $-2.78\times10^{-3}$ & $-2.09\times10^{-2}$ \\
        0.6   & $-1.93\times10^{-3}$ & $0$ & $-1.93\times10^{-3}$ & $-1.59\times10^{-2}$ \\
        0.7   & $-4.31\times10^{-4}$ & $0$ & $-4.31\times10^{-4}$ & $-5.10\times10^{-3}$ \\
        0.9   & $-2.08\times10^{-6}$ & $0$ & $-2.08\times10^{-6}$ & $-9.31\times10^{-5}$ \\
        \bottomrule
    \end{tabular}
\end{table}

Table~\ref{tab:5} gives the main results. It can be seen that the proton term is always negative ($\mathcal E_p<0$) and the electron term is always positive ($\mathcal E_e>0$); the competition between the two determines whether the black hole mass increases or decreases: in weak and intermediate magnetic fields ($BM\lesssim0.551$), the polar region is still open within the range of radii occupied by particles, and what is mainly captured is positive-energy electrons; $\mathcal E_e$ is two to five orders of magnitude larger than $|\mathcal E_p|$, so the net energy change of the black hole is positive and the mass increases. This conclusion is qualitatively consistent with the result of Ref.~\cite{5} under a spherically symmetric density. As the magnetic field increases, $\theta_c$ shrinks on the horizon (Fig.~\ref{fig:1}), and more importantly, it collapses at a finite radius (Table~\ref{tab:2}), and the polar region disappears first in the region where the particles actually reside (Fig.~\ref{fig:2}), so the particles pushed toward the horizon change from electrons to protons. Consequently $\mathcal E_e$ decays rapidly and becomes strictly zero at $BM\simeq0.569$---at this point the polar region itself has completely collapsed, and there is no longer any region in the entire capture domain where electrons are accelerated inward. At the same time, the capture of negative-energy protons remains appreciable, and finally at
\begin{equation}
BM\gtrsim0.551\quad (a/M=0.7)
\end{equation}
it exceeds the electron contribution, and both the net energy change and net angular momentum change of the black hole are negative, $\Delta\mathcal E<0$ and $\Delta\mathcal L<0$, so energy extraction can be realized even with a spherically symmetric density. This forms a sharp contrast with the Kerr black hole in the Wald solution: in the latter $\theta_c$ is fixed at $56.12^\circ$ and independent of the magnetic field, the polar region does not collapse with radius or magnetic field, and therefore special anisotropic densities are required to enhance the capture of negative-energy particles\cite{5}. For example, the anisotropic density they chose is\cite{5}
\begin{equation}
n(r, \theta) = \mathcal{X}(r)\Psi(\theta), \quad \mathcal{X}(r) = n_+(r_+/r)^2, \quad \Psi(\theta) = (1 - \cos\theta)^2.
\end{equation}
In the Kerr--Bertotti--Robinson black hole, the ``special conditions'' required to achieve energy extraction are provided by the external magnetic field itself, without the need to set a special density distribution.

\section{Conclusions}
In this paper, in the background of the Kerr--Bertotti--Robinson black hole, we have systematically studied the electromagnetic field structure of a rotating black hole immersed in an asymptotically uniform magnetic field, as well as the electrodynamic processes of charged particles. Unlike previous studies of Kerr black holes based on the Wald solution, the Kerr--Bertotti--Robinson black hole is a new class of exact magnetized black hole solution. It satisfies the Einstein--Maxwell equations, its Weyl tensor is algebraic type D, its electromagnetic field is non-aligned and nonzero, the magnetic field is asymptotically uniform at infinity, the ergosphere is bounded, and both axes can be regularized simultaneously by the conical parameter \(C\). Therefore, the Kerr--Bertotti--Robinson black hole provides a more suitable magnetized black hole background than models such as Kerr--Melvin.

Regarding the electromagnetic field structure, the critical angle \(\theta_c\) of the electromagnetic invariant \(\vec{E}\cdot\vec{B}=0\) of the Kerr--Bertotti--Robinson black hole varies with the magnetic field \(BM\), rather than being fixed to a single value as in the Kerr black hole in the Wald solution. In the polar region near the horizon, protons are accelerated outward and electrons inward; in the equatorial region, their roles are interchanged. As the magnetic field increases, the polar region shrinks and the equatorial region expands. When \(a/M=0.7,BM\approx 0.569\), the polar region at the horizon almost disappears. More importantly, the zero interface of \(\vec{E}\cdot\vec{B}=0\) is not a cone. It contracts with radius and completely disappears after \(r\gtrsim2.2517r_+\), so the acceleration direction of particles must be determined point by point.

We also pointed out that since this spacetime is asymptotically Bertotti--Robinson, \(A_\phi\) saturates at large distances instead of diverging as \(r^2\), and magnetic field lines are no longer cylindrical. The capture criterion ``\(r\sin\theta\le r_+\)'' in the Wald solution is no longer applicable. We instead used the gauge-invariant magnetic flux criterion \(A_\phi\in[A^{H}_{\min},A^{H}_{\max}]\) to determine the capture domain, and performed a two-dimensional integral on the \((r,\theta)\) plane by combining the ergosphere condition, the acceleration-direction condition, and the \(K_i\ge0\) condition.

Regarding particle energy and angular momentum, the Kerr--Bertotti--Robinson black hole gives a picture clearly different from that of the Kerr black hole in the Wald solution. Because of the different electromagnetic four-potential, protons carry negative energy and negative angular momentum, while electrons carry positive energy and positive angular momentum. Under weak magnetic fields, the polar region is still open within the capture domain, positive-energy electron injection dominates, and the black hole mass increases. When \(a/M=0.7\) and \(BM\gtrsim0.55\), the polar region almost collapses at the radii where the particles reside (complete collapse occurs at \(BM\approx0.569\)), and the capture of negative-energy protons becomes dominant instead. The net energy change and net angular momentum change of the black hole are negative, so energy extraction can be realized under the usual spherically symmetric density. This result shows that in the Kerr--Bertotti--Robinson black hole, the ``special conditions required to achieve energy extraction'' are provided by the strength of the external magnetic field itself, without introducing an anisotropic particle density distribution.

\noindent {\bf Acknowledgments}

\noindent
This work is supported by the National Natural Science Foundation of China (Grants Nos. 12375043, 12575069), Chongqing Normal University Fund Project (Grant No. 26XLB001), and Science and Technology Research Program of Chongqing Municipal Education Commission (Grant No. KJZD-K202600503).

\end{document}